\documentclass{ifacconf}

\usepackage{natbib}
\usepackage{amsmath,graphicx}
\usepackage{amssymb}
\usepackage[dvips]{epsfig}
\usepackage{subfigure}
\usepackage{multirow}
\usepackage{mathtools}

\newcounter{myAlgoCounter}
\usepackage{ssCorrections}
\xdefinecolor{colSS}{rgb}{1,0,0} %red

\def \pr {\mathsf{P}}
\def \E {\mathbf{E}}
\def \V {\mathbf{Var}}
\def \t {\boldsymbol{\theta}}
\def \s {\boldsymbol{\xi}}
\def \x {\mathbf{x}} 
\def \u {\mathbf{u}}
\def \y {\mathbf{y}}
\def \f {\mathbf{f}}
\def \h {\mathbf{h}}

\def \ep {\boldsymbol{\varepsilon}}

\begin{document}

\begin{frontmatter}

\title{A Probabilistic Approach to\\Robust Optimal Experiment Design with\\Chance Constraints}

\author[First]{Ali Mesbah} 
\author[Second]{Stefan Streif} 

\address[First]{Department of Chemical and Biomolecular Engineering,\\ University of California, Berkeley, USA; \\ e-mail: {\tt mesbah@berkeley.edu}.}
\address[Second]{Institute for Automation and Systems Engineering, Ilmenau University of Technology, Germany; \\e-mail: {\tt stefan.streif@tu-ilmenau.de}.}

%\address[First]{Massachusetts Institute of Technology, Cambridge, MA, USA}                         
%\address[Second]{Otto-von-Guericke University Magdeburg, Magdeburg, Germany}

%\begin{keyword}                           % Five to ten keywords,  
%Cicero; Catiline; orations.               % chosen from the IFAC 
%\end{keyword}                             % keyword list or with the 
                                          % help of the Automatica 
                                          % keyword wizard

\begin{abstract}                          % Abstract of not more than 250 words.
Accurate estimation of parameters is paramount in developing high-fidelity models for complex dynamical systems. Model-based optimal experiment design (OED) approaches enable systematic design of dynamic experiments to generate input-output data sets with high information content for parameter estimation. Standard OED approaches however face two challenges: (i) experiment design under incomplete system information due to unknown true parameters, which usually requires many iterations of OED; (ii) incapability of systematically accounting for the inherent uncertainties of complex systems, which can lead to diminished effectiveness of the designed optimal excitation signal as well as violation of system constraints. This paper presents a robust OED approach for nonlinear systems with arbitrarily-shaped time-invariant probabilistic uncertainties. Polynomial chaos is used for efficient uncertainty propagation. The distinct feature of the robust OED approach is the inclusion of chance constraints to ensure constraint satisfaction in a stochastic setting. The presented approach is demonstrated by optimal experimental design for the JAK-STAT5 signaling pathway that regulates various cellular processes in a biological cell.                     
\end{abstract}
        
\end{frontmatter}

\section{Introduction}

The prediction capability of first-principles models of complex dynamical systems largely relies on the accuracy of model parameters. Parameter estimation for complex systems is often a challenging task due to nonlinear nature of system dynamics as well as system uncertainties and disturbances that are ubiquitous in real-world applications. Hence, dynamic experiments that provide as much information as possible about the system dynamics in the face of system nonlinearities and uncertainties are crucial for obtaining accurate estimates of model parameters. This consideration has led to the development of model-based optimal experiment design approaches (e.g., see \citep{pro08} and the references therein) that facilitate systematic design of the system excitation inputs to maximize the information content of dynamic experiments. Optimal experiment design (OED) also enables seeking trade-offs between the economic costs and information content of dynamic experiments, which is particularly important when dynamic experiments are economically expensive \citep{bom06}.     

The primary challenge in model-based OED approaches arise from the fact that the OED problem depends on the unknown model parameters. Hence, the excitation inputs are designed based on the current best estimate of the parameters, which can be largely different from the true parameter values. Designing experiments under incomplete system information (i.e., unknown true parameters) is likely to diminish the effectiveness of the optimal excitation inputs and, therefore, lead to loss in information content of the designed experiments \citep{asp02}. Another difficulty in OED originates from the inherent system uncertainties that can result in violations of (state and/or output) constraints incorporated into the OED problem. 

One approach to deal with the inadequate system information and system uncertainties in OED is to adopt sequential experiment design strategies that repeatedly estimate the model parameters and redesign the experiments till parameter estimates with admissible uncertainty are obtained \citep{asp02}. Sequential OED can however be economically infeasible for certain applications due to high costs of experiments. Alternatively, robust OED approaches enable one to systematically account for the effects of uncertainty on the designed experiments by devising the excitation inputs based on a prespecified parameter range around the nominal values. Various robust OED formulations have been proposed in terms of max-min optimization problems, in which uncertainties are typically assumed to be deterministic and bounded \citep{pro85,kor04,fla06,goo07}. 
In max-min robust OED approaches, the excitation inputs are designed with respect to worst-case uncertainty realizations, and constraints are satisfied for all admissible values of uncertainties. Such robust OED approaches discard the statistical properties of uncertainties, and can be conservative if the worst-case uncertainty realizations have a small probability of occurrence.                

This paper considers the problem of robust OED for nonlinear systems with time-invariant probabilistic uncertainties. In the proposed approach, parametric and initial condition uncertainties are described by probability distributions (instead of bounded sets), which can often be readily obtained from a (previous) model identification procedure \citep{lju99}. The probabilistic experiment design framework circumvents the conservatism of worst-case OED approaches, as the probability of occurrence of different uncertainty realizations is directly accounted for in design of experiments. More importantly, chance (a.k.a.\ probabilistic) constraints are incorporated into the OED problem to seek a trade-off between the information content of a designed experiment and allowing for prespecified levels of (operational) risks during the experiment. Chance constraints enable satisfaction of constraints with a desired probability level in the presence of system uncertainties (e.g., see \citep{sch99,cal06,old13,mes14a} for the application of chance constraints in stochastic optimal control problems). To the best of the authors' knowledge, this paper is the first contribution of its kind that considers chances constraints for OED of nonlinear systems in a probabilistic setting.     

A nonlinear optimization problem with chance constraints is presented for robust OED (Section~\ref{sec:PF}). The objective function of the OED problem is defined in terms of the weighted sum of the expected value and variance of a scalar metric of the Fisher information matrix (Section~\ref{sec:PF}). Such an OED objective function enables maximizing the information content of the dynamic experiments, while minimizing the variance of the information content with respect to realizations of the probabilistic system uncertainties. Efficient propagation of uncertainties through the nonlinear system dynamics poses a challenge in evaluating the objective function. The generalized polynomial chaos (PC) framework \citep{wie38,xiu02} is used as a computationally efficient spectral tool for probabilistic uncertainty propagation (Section \ref{sec:PCE}). The PC framework replaces the implicit mappings between the uncertain system variables/parameters and dynamic state variables with a series of orthogonal polynomials, whose statistical moments can be readily computed from the expansion coefficients (e.g., see \citep{fis09,fag12,mes14b,pau14}, and references therein for applications of PC expansions). The Cantelli-Chebyshev inequality \citep{mar79} is used to convert chance constraints into deterministic expressions and, therefore, obtain a computationally tractable optimization problem (Section \ref{sec:CC}). The proposed robust OED approach is demonstrated for the JAK-STAT5 signaling pathway \citep{ber12} in a biological cell with probabilistic parametric uncertainties (Section \ref{sec:example}).                    

%%%%%%%%%%%%%%%%%%%%%%%%%%%%%%%%%%%%%%%
\section{Problem Formulation}
\label{sec:PF}

Consider a continuous-time, uncertain nonlinear system 
\begin{subequations} \label{e_sys}
\begin{align}
& \dot{\x}(t)  = \f(\x(t),\u(t),\t), \quad \x(0) = \x_0  \label{e_1} \\
& \y(t)  = \h(\x(t)) + \ep(t),   \label{e_2} 
\end{align}   
\end{subequations} 
where $t \in [0,t_f]$ denotes the time; $\x \in \mathbb{R}^{n_x}$, $\u \in \mathbb{R}^{n_u}$, and $\y \in \mathbb{R}^{n_y}$ denote the system states, (excitation) inputs, and outputs, respectively; $\t \in \mathbb{R}^{n_{\theta}}$ denotes the time-invariant uncertain system parameters with known probability distribution functions (PDFs) $\{\pr_{\theta_i}\}_{i=1}^{n_\theta}$; $\x_0$ denotes the initial states that are considered to be uncertain with known PDFs $\{\pr_{x_{0,i}}\}_{i=1}^{n_x}$; $\f:\mathbb{R}^{n_x} \times \mathbb{R}^{n_u} \times \mathbb{R}^{n_\theta} \rightarrow \mathbb{R}^{n_x}$ and $\h:\mathbb{R}^{n_x} \rightarrow \mathbb{R}^{n_y}$ denote the nonlinear system and (possibly nonlinear) model output functions, respectively, which are typically represented by a set of differential algebraic equations; and $\ep \in \mathbb{R}^{n_y}$  denotes zero-mean additive measurement noise that has a known variance-covariance matrix $\Sigma \in \mathbb{R}^{n_y\times n_y}$. Define a probability space $(\Omega,\mathcal{F},\pr)$ on the basis of the sample space $\Omega$, $\sigma$-algebra $\mathcal{F}$, and the probability measure $\pr$ on $\Omega$. The time-invariant probabilistic uncertainties $[\x_0^\top \; \t^\top] \in \mathbb{R}^{n_\xi}$ (with $n_\xi \le n_\theta + n_x$) are functions of standard random variables $\s \coloneqq [\xi_1,\ldots,\xi_{n_\xi}]^\top$ with known independent PDFs $\{\pr_{\xi_i}\}_{i=0}^{n_\xi}$ over the common support $\Omega$. Note that $\xi_i \in \mathcal{L}^2(\Omega,\mathcal{F},\pr)$, where $\mathcal{L}^2(\Omega,\mathcal{F},\pr)$ is the Hilbert space of all random variables with finite variance $\E[\xi_i^2]<\infty$.       

This paper considers the problem of robust OED for parameter estimation.\footnote{The problem of robust OED for model discrimination \textit{without} chance constraints is addressed in our paper \citep{str14}.} The information content of experiments can be quantified in terms of some scalar metric of the Fisher information (FI) matrix $F(t_f)$ defined by \citep{bar74}
\begin{equation} \label{e_FIM}
F(t_f) = \int_0^{t_f}  \left(\frac{\partial \h}{\partial \x} \frac{\partial \x (t)}{\partial \t} \right)^\top \Sigma^{-1} \left(\frac{\partial \h}{\partial \x}\frac{\partial \x (t)}{\partial \t}\right)   dt,
\end{equation}   
where sensitivities $\frac{\partial \x (t)}{\partial \t} $ are obtained through integrating
\begin{equation*} 
\frac{d}{dt}\frac{\partial \x}{\partial \t}(t) = \frac{\partial \f}{\partial \x}\frac{\partial \x}{\partial \t}(t) + \frac{\partial \f}{\partial \t}, \quad \frac{\partial \x}{\partial \t}(0) = \frac{\partial \x_0}{\partial \t}. 
\end{equation*}   
The FI matrix describes the amount of information that system outputs $\y$ provide on the unknown parameters $\t$. The FI matrix accounts for the effects of measurement noise $\ep(t)$ and sensitivities of the system states to variations in the model parameters (i.e., $\frac{\partial \x (t)}{\partial \t} $). Under the assumption of unbiased parameter estimates and uncorrelated measurement noise, the inverse of the Fisher information matrix $F(t_f)$ provides an approximation of the Cram\'{e}r-Rao lower bound \citep{bar74}, which is closely related to the lower bound of variance-covariance matrix of the estimated parameters.     

The E-optimality criterion \citep{pro08} is adopted as a scalar metric of the FI matrix to formulate the OED problem.\footnote{The proposed robust OED approach can be straightforwardly adapted for other optimality criteria such as A- and D-optimal designs.} The E-optimality criterion aims to maximize the minimum eigenvalue of the Fisher information matrix, i.e.,
\begin{equation*} 
\Phi(F(t_f)) \coloneqq \max \big[\lambda_{min}\big(F(t_f)\big) \big].
\end{equation*}   
Hence, E-optimal designs in effect minimize the length of the largest uncertainty axis of the joint confidence region of parameters, which corresponds to the largest parameter errors. 

The primary challenge in performing OED results from the fact that the optimality criterion depends on the current estimates of the to-be-estimated parameters (required to evaluate the FI matrix~\eqref{e_FIM}). The uncertainty in initial estimates of the to-be-estimated parameters can render the OED in practice largely ineffective, as the excitation inputs are designed on the basis of an inadequate description of the system dynamics. In addition, plant-model mismatch due to initial condition and parametric uncertainties is likely to further diminish the effectiveness of the designed excitation inputs.           

This paper proposes the following robust OED formulation for the nonlinear system~\eqref{e_sys} to systematically incorporate the knowledge of time-invariant probabilistic uncertainties into the OED problem.

\textbf{Problem 1 (Robust optimal experiment design with chance constraints)}: The optimal excitation inputs $\u^\ast$ to system~\eqref{e_sys} that maximize the information content of dynamic experiments, while being robust to the probabilistic uncertainties in $[\x_0^\top \; \t^\top]$, are defined by   
\begin{equation}
\label{e_P1}
\arraycolsep=0pt
\begin{array}{rclr}
\multicolumn{4}{l}{\u^\ast \coloneqq \underset{\u}{\arg \min}\quad  \E [\Phi(F(t_f))] + w \V [\Phi(F(t_f))] }\\[1eM]
\multicolumn{2}{r}{\text{subject to :}}\\[0.6eM]
\dot{\x}(t)  &=& \f(\x(t),\u(t),\t),  & t \in [0,t_f] \\[0.8eM]
\frac{d}{dt}\frac{\partial \x}{\partial \t}(t) &=& \frac{\partial \f}{\partial \x}\frac{\partial \x}{\partial \t}(t) + \frac{\partial \f}{\partial \t},  & t \in [0,t_f] \\[0.8eM]
\frac{d}{dt}F(t) &=& \left( \frac{\partial \h}{\partial \x}\frac{\partial \x (t)}{\partial \t}\right)^{\!\!\top}\!\!\Sigma^{-1}\!\!\left(\frac{\partial \h}{\partial \x}\frac{\partial \x (t)}{\partial \t}\right),  & t \in [0,t_f] \\[1eM] \multicolumn{3}{l}{\mathbf{Pr}[b_ix_i(t) \geq x^{\text{max}}_i  ] \leq \beta_i, \forall i \in \mathcal{I},}  &\ t \in [0,t_f] \\[0.6eM]
\u(t) &\in& \mathbb{U},& t \in [0,t_f] \\ 
F(0)&=&0 \\
\frac{\partial \x}{\partial \t}(0) &=& \frac{\partial \x_0}{\partial \t} \\ 
x_i(0) &\sim& \pr_{x_{0,i}}, & \hspace{-2em}i=1,\ldots,n_x  \\ 
\theta_i &\sim& \pr_{\theta_{i}},  & \hspace{-2em}i=1,\ldots,n_{\theta},  
\end{array}
\end{equation} 
where $\E[\cdot]$ and $\V[\cdot]$ denote the expected value and variance of a stochastic variable; $w$ denotes a scalar weight function; $\mathbf{Pr}$ denotes probability; the scalar values $b_i \in \mathbb{R}$ and $x^{\text{max}}_i \in \mathbb{R}$ define the state constraints; $\beta_i \in (0, \; 1) \subset \mathbb{R}$ denotes the lower bound of the desired probability that each state constraint should satisfy in the probabilistic uncertainty setting; $\mathcal{I} \subseteq \{1,\ldots, n_x\}$ denotes a subset of states for which the chance constraints are defined; and $\mathbb{U}$ denotes the convex compact set of input constraints. Note that the initial conditions $\{x_i(0)\}_{i=1}^{n_x}$ and parameters $\{\theta_i\}_{i=1}^{n_\theta}$ have probability distributions $\pr_{x_{0,i}}$ and $\pr_{\theta_{i}}$, respectively.         

The robust OED problem~\eqref{e_P1} can effectively account for the uncertainty in the initial parameter estimates and the inherent system uncertainties. This is due to (i) defining the objective function in terms of the statistical moments of the metric $\Phi(F(t_f))$ that are evaluated with respect to probabilistic uncertainty realizations\footnote{This is in contrast to classical OED approaches, which evaluate a metric of the FI matrix merely at one realization of uncertainties.}, and (ii) including chance constraints to seek trade-offs between maximizing the information content of the experiments and constraint satisfaction (typically associated with operational risks) in a stochastic setting. The variance term in the objective function of~\eqref{e_P1} enables minimizing variations in the information content of the experiments due to probabilistic uncertainties.

The key challenges in solving Problem 1 are to efficiently propagate the probabilistic uncertainties through the nonlinear system dynamics \eqref{e_sys} and to obtain a computationally tractable surrogate for the chance constraints in \eqref{e_P1}. To this end, polynomial chaos is introduced next, which provides a computationally efficient means to evaluate statistical moments of a stochastic variable.  

\section{Polynomial Chaos for Uncertainty Propagation}
\label{sec:PCE}

This work adopts the generalized polynomial chaos framework \citep{wie38,xiu02} to efficiently propagate the time-invariant probabilistic uncertainties $[\x_0^\top \; \t^\top]$ through the nonlinear system \eqref{e_sys}. In the PC framework, a second-order stochastic variable $\psi(\s) \in \mathcal{L}^2(\Omega,\mathcal{F},\pr)$ is defined in terms of an expansion of orthogonal polynomial basis functions
\begin{align} \label{e_PCE}
\psi(\s) =  \sum\limits_{k=0}^{\infty}a_k\Phi_{k}(\s) ,  
\end{align}
where $a_k$ denotes the expansion coefficients, and $\Phi_{k}$ denotes polynomial basis functions of maximum degree $m$ with respect to the random variables $\s$. The basis functions belong to the Askey scheme of polynomials, which entails a set of orthogonal basis functions in the Hilbert space defined on the support of the random variables \citep{xiu02}. Hence, $\langle \Phi_i(\s), \Phi_j(\s) \rangle = \langle \Phi_i^2(\s) \rangle\delta_{ij}$, where $\langle h(\s), g(\s) \rangle=\int_{\Omega}h(\s)g(\s)\pr_{\s}d\s$ denotes the inner product induced by $\pr_{\s}$, and $\delta_{ij}$ denotes the Kronecker delta function. The basis functions $\Phi_{k}$ are chosen in accordance with the PDFs of the uncertain variables $\s$. The truncated form of~\eqref{e_PCE} used in practice takes the the form   
\begin{align} \label{e_PCE_T}
\hat{\psi}(\s) \coloneqq  \sum\limits_{k=0}^{L}a_k\Phi_{k}(\s) = \mathbf{a}^\top\boldsymbol{\Phi}(\s)  
\end{align}
with $L+1=\frac{(n_\xi+m)!}{n_\xi!m!}$ being the total number of terms in the expansion; $\mathbf{a} \coloneqq [a_0, \ldots, a_L]^{\top}$; and $\boldsymbol{\Phi}(\s) \coloneqq [\Phi_{0}(\s), \ldots, \Phi_{L}(\s)]^\top$. Owing to the orthogonality property of the polynomial basis functions, the statistical moments of $\hat{\psi}$ can be computed merely based on the expansion coefficients $\mathbf{a}$ in a computationally efficient manner (e.g., see \citep{fis09}).

The probabilistic collocation method (see discussion and references in \citep{fag12}) is used to determine the coefficients $\mathbf{a}$ in \eqref{e_PCE_T}. The collocation method requires the residuals 
\begin{equation*} 
R(\mathbf{a},\s) = \hat{\psi}(\s) - \psi(\s)
\end{equation*} 
be orthogonal to each basis function $\Phi_{k}$  
\begin{equation} \label{e_PCM}
\int_{\Omega} R(\mathbf{a},\s) \Phi_{k}(\s) d\pr_{\s} =0, \quad k=0,\ldots,L.  
\end{equation} 
Provided that the basis functions are non-zero terms, the coefficients $\mathbf{a}$ can be estimated by computing the residuals $R(\mathbf{a},\s)$ at $n_c$ samples (i.e., collocation points) of random variables $\s$ with non-zero probability $\pr_{\s}$ \citep{tat97}. Alternatively, the expansion coefficients $\mathbf{a}$ in \eqref{e_PCE_T} can be determined using the Galerkin-projection method (see \citep{gha91}).      

\section{Deterministic Surrogate for Chance Constraints }
\label{sec:CC}

To solve the robust OED Problem 1, the chance constraints in \eqref{e_P1} should be replaced with deterministic expressions. The Cantelli-Chebyshev inequality is used to obtain a computationally tractable optimization problem.\footnote{The Cantelli-Chebyshev inequality has also been used for converting chance constraints in the context of stochastic predictive control \citep{far13}. }   

\textbf{Theorem 1 (Cantelli-Chebyshev inequality \citep{mar79}):} Let $\psi$ be a stochastic variable with a finite second-order moment. Then,  
\begin{align} \label{e_Cant}
\mathbf{Pr} [ \psi - \E[\psi] \geq \alpha ] \leq \frac{\V[\psi]}{\V[\psi] + \alpha^2} , \quad \forall \alpha \in \mathbb{R}_0^{+}.
\end{align}
\hfill$\blacksquare$ 

Consider the chance constraints
\begin{align} \label{e_CC_O}
\mathbf{Pr}[b_ix_i(t) \geq x^{\text{max}}_i  ] \leq \beta_i, \quad \forall i \in \mathcal{I}
\end{align}
in \eqref{e_P1}, which are independently defined in terms of state variables $x_i(t)$. Define $\delta x_i \geq 0$ such that 
\begin{equation} \label{e_CC_T1} 
b_i\E[x_i(t)] + \delta x_i \leq x^{\text{max}}_i.
\end{equation}  
The chance constraints~\eqref{e_CC_O} then satisfy
\begin{align*} 
\mathbf{Pr}[b_ix_i(t) \geq x^{\text{max}}_i  ] \leq \mathbf{Pr}[ b_ix_i(t) \geq b_i\E[x_i(t)] + \delta x_i ]. 
\end{align*}
Since the states $\{x_i(t)\}_{i=1}^{i=n_x}$ are stochastic variables due to the probabilistic time-invariant uncertainties in~\eqref{e_sys}, the Cantelli-Chebyshev inequality in Theorem 1 implies that 
\begin{align*} %\label{e_CC_T}
\mathbf{Pr}[ b_ix_i(t) \geq b_i\E[x_i(t)] + \delta x_i ] \leq \frac{b_i^2 \V[x_i(t)]}{b_i^2 \V[x_i(t)] + \delta x_i^2}, 
\end{align*}
where the fulfillment of chance constraints~\eqref{e_CC_O} requires   
\begin{align} \label{e_CC_T2}
\frac{b_i^2 \V[x_i(t)]}{b_i^2 \V[x_i(t)] + \delta x_i^2} \leq \beta_i. 
\end{align} 
Hence, replacing~\eqref{e_CC_T2} in~\eqref{e_CC_T1} for $\delta x_i$ and rearranging the resulting inequality will lead to a deterministic surrogate for each individual chance constraint in \eqref{e_CC_O} 
\begin{align} \label{e_CC}
b_i\E[x_i(t)] \leq x^{\text{max}}_i - \sqrt{b_i^2 \V[x_i(t)]}\sqrt{\frac{1-\beta_i}{\beta_i}}. 
\end{align}

In this work, the stochastic state variables $\{x_i(t)\}_{i=1}^{i=n_x}$ and their moments are approximated using polynomial chaos expansions (see \eqref{e_PCE_T}). Thus, $x_i(t)$ is replaced with $\hat{x}_i(t)$ in the deterministic constraint~\eqref{e_CC} and the first- and the second-order moments of $\hat{x}_i(t)$ are given by (see, e.g., \citep{fis09})
\begin{equation*} %\label{e6}
\E[\hat{x}_i(t)]=a_0
\end{equation*}  
\begin{equation*} %\label{e7}
\V[\hat{x}_i(t)] = \sum^{L}_{k=1} a_k^2 \E[\Phi^2_{k}(\s)].
\end{equation*}     
Note that the terms $\{ \E[\Phi^2_{k}(\s)] \}_{k=1}^L$ are computed only once, prior to performing the OED.      

\section{Robust Optimal Experiment Design for the JAK-STAT5 Cell-signaling Pathway}
\label{sec:example}

The dynamics of complex biological systems such as metabolic and cell-signaling pathways in living cells are often described by nonlinear differential equations, which typically have several kinetic parameters. The fidelity of these dynamic models is largely dependent on the quality of the estimated parameters. This work considers the problem of OED in the presence probabilistic uncertainties for the JAK-STAT5 cell-signaling pathway, which is a fast-track signal transduction pathway for transferring information from cell-surface receptor into the nucleus. Deregulation of the STAT5 signaling pathway is shown to be connected to human cancer \citep{ber12}.       

The STAT5 cell-signaling mechanism entails phosphorylation of the STAT5 molecules, which is governed by the EPO receptor on the cell membrane. The activated STAT5 molecules undergo a dimerization, so that STAT5 dimers can enter the cell nucleus to trigger the transcription of target genes. The STAT5 molecules are then dephosphorylated through separation of dimers, and the single STAT5 molecules re-enter the cytoplasm \citep{pei07}. Assuming that no concentration gradient occurs in the cell due to fast transport mechanisms from the cell membrane to nucleus, the STAT5 cell-signaling dynamics can be described by \citep{pei07}
\begin{equation} \label{e_LC}
\begin{array}{ll}
\dot{x}_1 & = -k_1x_1u(t) + k_2x_3(t-\tau) \\
\dot{x}_2 & = -k_3x_2^2 + k_1x_1u(t) \\
\dot{x}_3 & = -k_2x_3 + k_3x^2_2 \\
\dot{x}_4 & = - k_2x_3(t-\tau) + k_2x_3.
\end{array}
\end{equation} 
In~\eqref{e_LC}, $x_1$, $x_2$, $x_3$, and $x_4$ denote the concentration of the unphosphorylated STAT5, activated STAT5, STAT5 dimer, and STAT5 molecules in the nucleus (mole fractions); $u$ denotes the EPO receptor activity (mole fraction); $k_1 \sim \boldsymbol{\beta}(2,5,1.90,2.34)$, $k_2 \sim \boldsymbol{\beta}(2,5,0.094,0.124)$, and $k_3 = 1.0$ denote rate constants (min$^{-1}$), with $\boldsymbol{\beta}$ being the four-parameter Beta distribution; and $\tau = 8$ denotes the delay parameter (min). The total amount of activated STAT5 and the total amount of STAT5 in the cytoplasm are defined by
\begin{align*} 
y_1 & = s_1(x_2+x_3) \\
y_2 & = s_2(x_1+x_2+x_3),
\end{align*} 
respectively, where $s_1=0.33$ and $s_2=0.26$ denote scaling constants. The model outputs $y_1$ and $y_2$ are assumed to be subject to zero-mean Gaussian noise with a $10\%$ variance. The delay element in~\eqref{e_LC} is approximated by adopting a delay chain approach \citep{pei07}.

\begin{table}[b!]
\caption{Average and maximum relative estimation errors of the model parameters with respect to the true parameter values}
%\vspace*{-0.10cm}
\begin{center}
\resizebox{7.5cm}{!}{%
\begin{tabular}{lcc|cc}
 \hline  \hline
& \multicolumn{2}{c}{Average Relative Error} &\multicolumn{2}{c}{Maximum Relative Error} \\  [0.5ex]  \hline 
 & $k_1$ & $k_2$ & $k_1$ & $k_2$ \\  \hline
OED & $1.68e^{-4}$ & $1.88e^{-5}$ & $6.84e^{-4}$ & $1.11e^{-4}$ \\
Robust OED & $5.88e^{-5}$ & $1.76e^{-5}$  & $3.15e^{-4}$ & $8.79e^{-5}$ \\ [1.0ex]  \hline
\end{tabular}  }
\end{center}
\label{T1}
\end{table} 

The STAT5 cell-signaling mechanism described by~\eqref{e_LC} comprises a nonlinear system with probabilistic time-invariant parametric uncertainties. To generate input-output data for estimating the kinetic parameters $k_1$ and $k_2$ with known PDFs, the robust OED problem~\eqref{e_P1} is adopted to design the EPO receptor activity profile (i.e., the excitation input). The polynomial chaos framework with the $4^{th}$-order Jacobi polynomial basis functions is used to propagate the Beta PDFs of $k_1$ and $k_2$ through the nonlinear system~\eqref{e_LC} (see Section~\ref{sec:PCE}). To ensure that the total amount of STAT5 in the cytoplasm ($y_2$) remains above a desired threshold at all times during the dynamic experiments in the presence of probabilistic uncertainties, the following chance constraint
\begin{equation} \label{e_Exp_CC}
\mathbf{Pr} [ y_2(t) \leq 0.038 ] \leq 0.05
\end{equation} 
is incorporated into the robust OED problem. The procedure in Section~\ref{sec:CC} is used to obtain a deterministic surrogate (see \eqref{e_CC}) for the above chance constraint. In the optimization problem, the excitation input (i.e., $u(t)$) is parameterized in a piecewise-constant manner with five equidistant intervals over the time horizon $[0, \; t_f]$, where $t_f=40$ min.           

\begin{figure}[t!] 
\centering
\subfigure[Parameter $k_1$]{
\includegraphics[width=225pt]{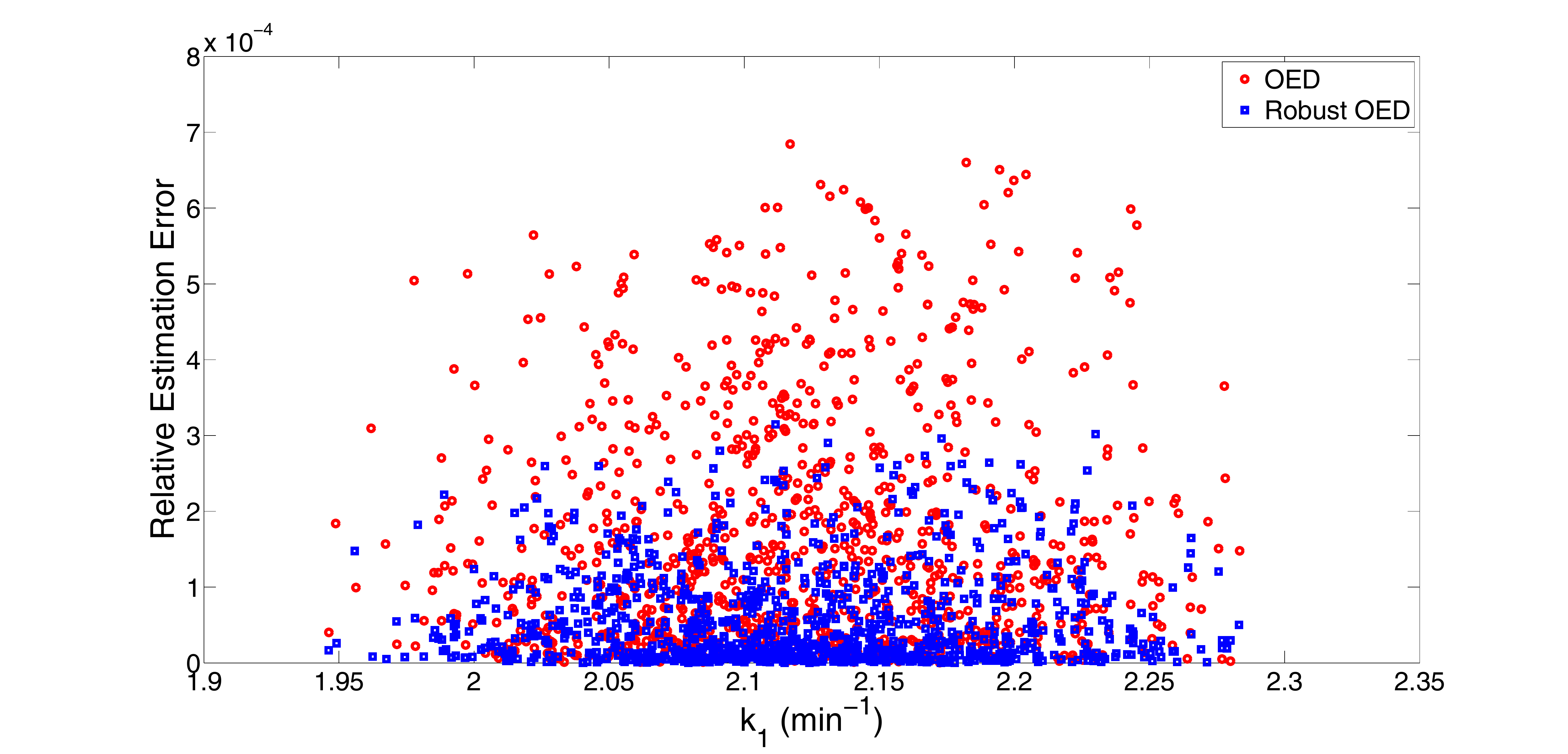}
}
\subfigure[Parameter $k_2$]{
\includegraphics[width=225pt]{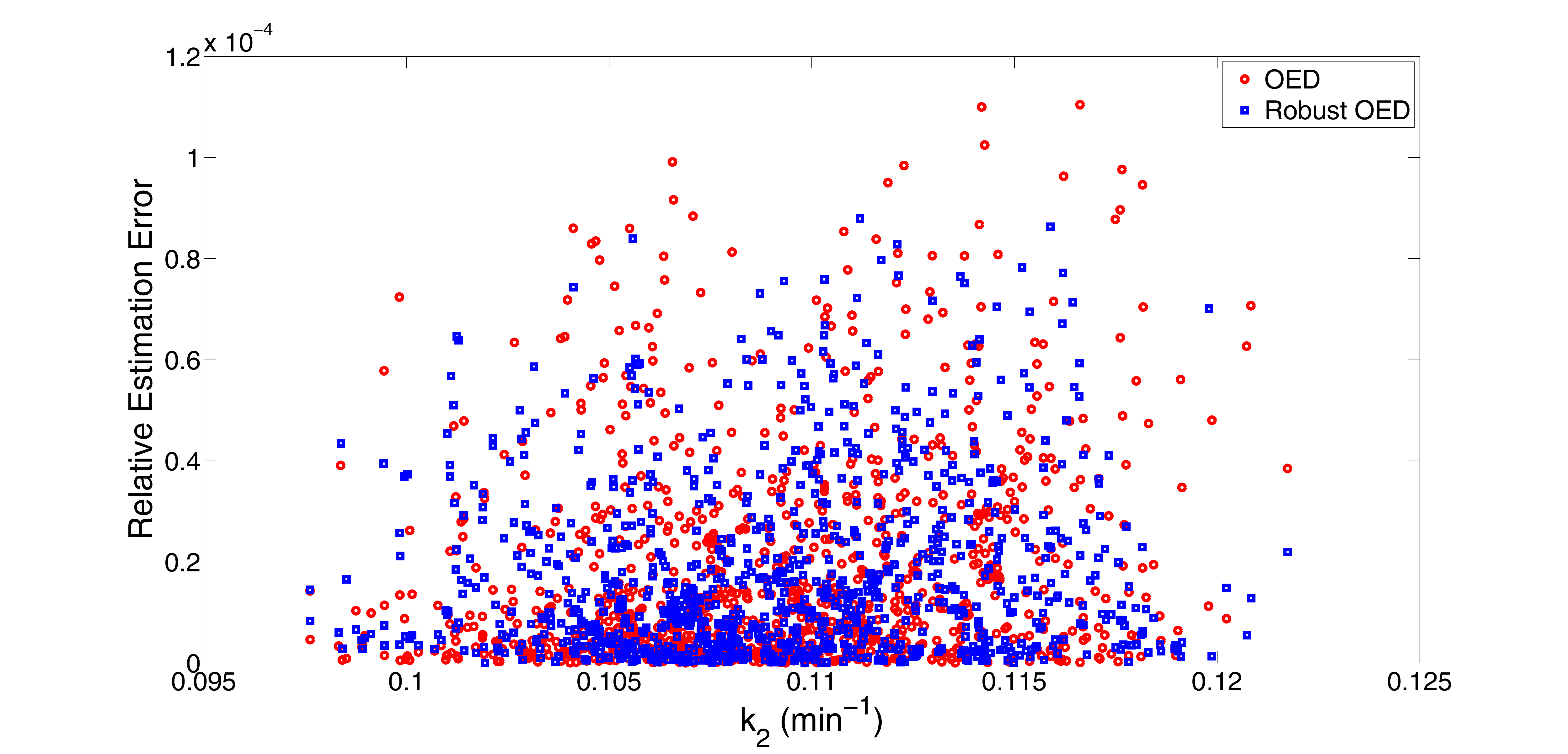}
}
\caption{Relative estimation errors of the model parameters with respect to the true parameter values. The parameter estimates are obtained by exciting the stochastic JAK-STAT5 cell-signaling pathway $1000$ times with the excitation inputs designed by the robust and standard OED approaches, and using the generated input-output data sets for parameter estimation.}
\label{fig1}
\end{figure} 

Monte Carlo simulations are performed to evaluate the performance of the proposed robust OED approach in dealing with the probabilistic system uncertainties. The designed excitation input is applied to the STAT5 cell-signaling mechanism \eqref{e_LC} in 1000 runs with different realizations of parametric uncertainties sampled from the known PDFs of $k_1$ and $k_2$. The generated input-output data is subsequently used for obtaining estimates for the parameters $k_1$ and $k_2$ using the weighted least-squares estimation method (see \citep{bar74}). The same procedure is also done for an excitation input designed based on a standard OED approach, against which the performance of the robust OED approach is compared. The standard OED approach does not take into account the statistical distributions of the unknown model parameters, and merely assumes some initial estimates for the parameters to perform the OED. Identical realizations of probabilistic uncertainties are used to compare the two OED approaches. 

\begin{figure}[b!] 
\centering
\subfigure[Standard optimal experiment design]{
\includegraphics[width=225pt]{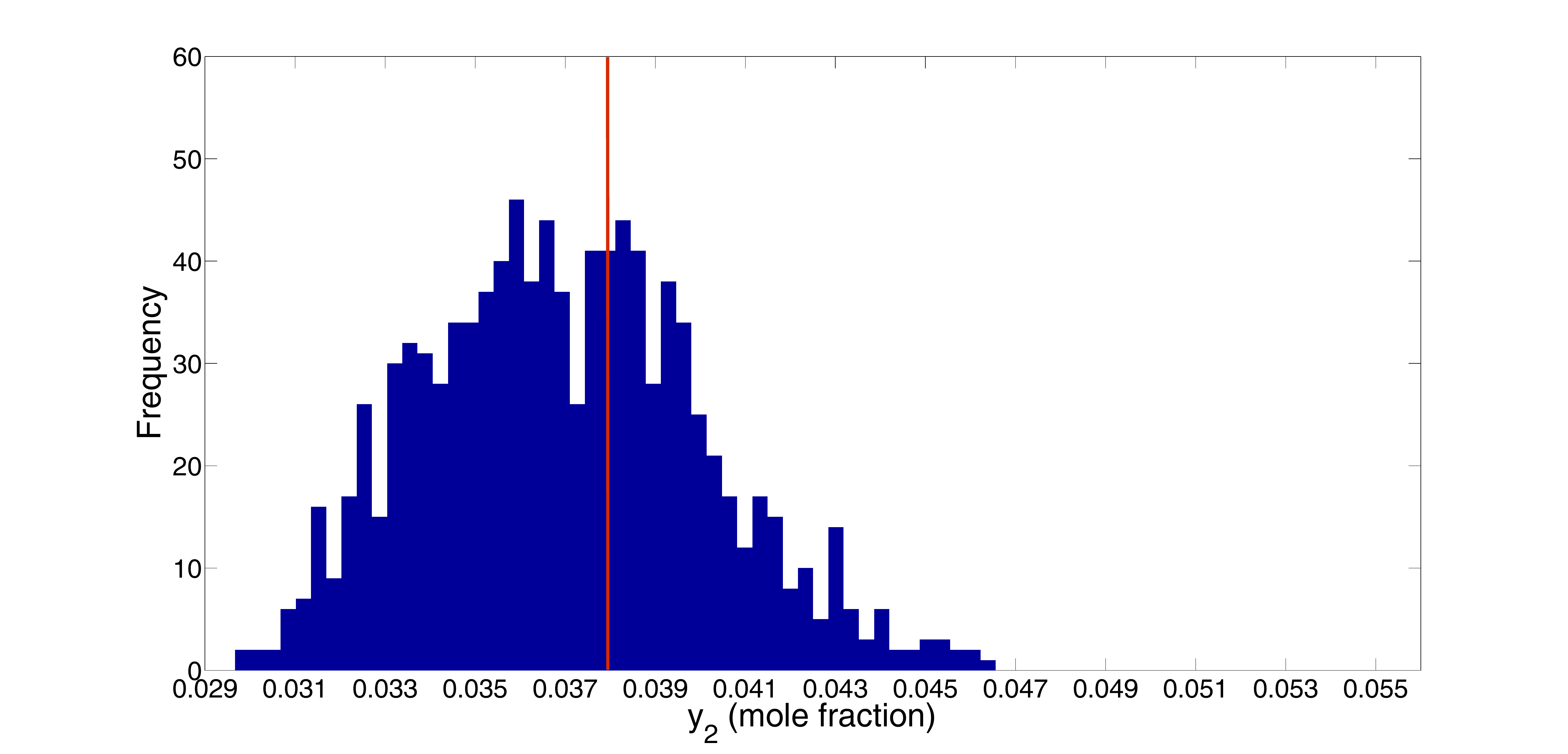}
}
\subfigure[Robust optimal experiment design]{
\includegraphics[width=225pt]{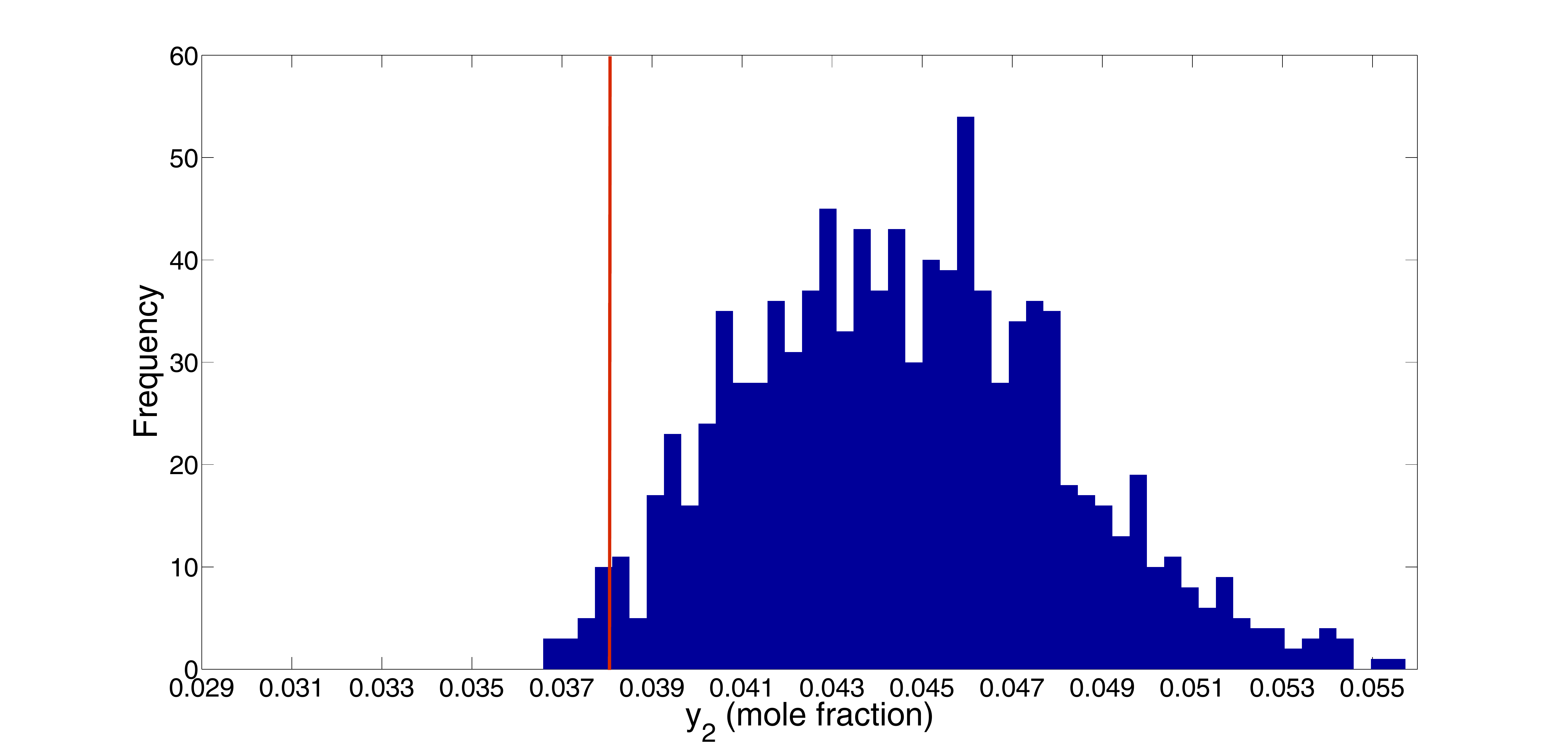}
}
\caption{Histograms of $y_2$ (the total amount of STAT5 in the cytoplasm) at time $40$ min. The red line represents the minimum admissible threshold (i.e., output constraint) for the total amount of STAT5 in the cytoplasm. The robust OED approach ensures constraint satisfaction in nearly $98$ $\%$ of signaling pathway excitations in the presence of probabilistic uncertainties, whereas the standard OED approach leads to approximately $46$ $\%$ constraint satisfaction.}
\label{fig2}
\end{figure} 

Figure~\ref{fig1} shows the relative parameter estimation errors computed with respect to the true parameter values in each Monte Carlo run. The robust OED approach results in smaller estimation errors for the parameter $k_1$ (see Figure~\ref{fig1}a). Table~\ref{T1} indicates that the average estimation error (computed over the 1000 Monte Carlo runs) for $k_1$ in the case of the standard OED approach is almost $3$ times larger than that in the case of the robust OED approach. More accurate parameter estimates are also obtained for $k_2$ (see Figure~\ref{fig1}b and Table~\ref{T1}), as the maximum estimation error for $k_2$ is lower when the system is excited with the input designed by the robust OED approach. Note that the OED problem~\eqref{e_P1} enables seeking systematic trade-offs between maximizing the information content of the dynamic experiments (through minimizing the expected value of some scalar metric of the FI matrix) and minimizing the variance of the chosen scalar metric of the FI matrix in the presence of probabilistic system uncertainties. This will be particularly useful when the accuracy of the to-be-estimated parameters should be traded-off against obtaining more consistent parameter estimates (with minimized variance) in a stochastic setting. In the presented simulation study, the weight function $w$ in~\eqref{e_P1} is selected to be small in order to generate an input-output data set with high information content, which in turn will lead to more accurate parameter estimates (i.e., small estimation errors in Figure~\ref{fig1}).

To demonstrate the ability of the robust OED approach in fulfilling the system constraints in a probabilistic sense (see \eqref{e_Exp_CC}), the histograms of $y_2$ at time $40$ min are shown in Figure~\ref{fig2}. The histograms are obtained based on $1000$ simulations of the stochastic JAK-STAT5 cell-signaling pathway when the receptor activity $u(t)$ is excited with the excitation input designed by the robust and standard OED approaches. The histograms of $y_2$ are shown for time $40$ min only, as the total amount of STAT5 in the cytoplasm is closest to its minimum admissible threshold value of $0.038$ at $t_f=40$ min. Figure~\ref{fig2} shows that the constraint on $y_2$ is satisfied in nearly $98\%$ of the Monte Carlo runs, which is greater than the lower bound (i.e., $95\%$) of the constraint satisfaction probability in~\eqref{e_Exp_CC}. On the other hand, the excitation input designed using the standard OED approach results in merely $46\%$ constraint satisfaction in the presence of probabilistic uncertainties of the system~\eqref{e_LC}. In general, the ability to guarantee state (output) constraint satisfaction in a stochastic setting is paramount in many OED applications to ensure safe, reliable, and high-performance system operation during dynamic experiments.    

\section{Conclusions}
\label{sec:Conclusions}

This paper presents a robust approach for optimal experiment design for nonlinear systems with arbitrarily-shaped probabilistic uncertainties. Polynomial chaos is used for efficient uncertainty propagation, and chance constraints are incorporated into the input design problem to ensure constraint satisfaction in a stochastic setting. The paper seems to be the first contribution of its kind that considers chance constraints for optimal experiment design. The simulation results for a cell-signaling pathway demonstrate the capability of the proposed approach in dealing with probabilistic system uncertainties and fulfilling system constraints in a probabilistic sense.

\bibliographystyle{plainnat}      
\bibliography{Literature_list}           
                                 
\end{document}